\documentclass{svmult}

\usepackage{epsfig}
\usepackage{amsmath}
\usepackage{natbib}

\def\Rreal{{\rm I\kern-.2em R}}

\font\tenbi=cmmib10
\newfam\bifam \def\bi{\fam\bifam\tenbi} \textfont\bifam=\tenbi

\def\vecv{{\textstyle{\bi v}}}

\newcommand{\apj}{    {\rm Astrophys. J.}\ }
\newcommand{\apjl}{   {\rm Astrophys. J. Lett.}\ }

\newcommand{\nat}{    {\rm Nature}\ }

\newcommand{\solphys}{{\rm Solar Phys.}\ }

\newcommand{\ssr}{    {\rm Space Sci. Rev.}\ }

\usepackage{amsfonts}   
\begin{document}
\title*{Reconstruction of  Solar Subsurfaces by Local Helioseismology}
\toctitle{Reconstruction of Solar Subsurfaces by Local Helioseismology}
%
%
\titlerunning{Reconstruction of Solar Subsurfaces}
%

\author{Alexander G. Kosovichev \& Junwei Zhao}
\authorrunning{Alexander G. Kosovichev \& Junwei Zhao}

\institute{New Jersey Institute of Technology, Newark, NJ 07103, USA;\\Stanford University, Stanford, CA 95305, USA\\E-mail: \texttt{sasha@bbso.njit.edu}}


\maketitle              

\begin{abstract}
\noindent 
Local helioseismology has opened new frontiers in our quest for understanding of the internal dynamics and dynamo on the Sun. Local helioseismology reconstructs subsurface structures and flows by extracting  coherent signals of acoustic waves traveling through the interior and  carrying information about subsurface perturbations and flows, from stochastic oscillations observed on the surface. The initial analysis of the subsurface flow maps reconstructed from the five years of  SDO/HMI data by time-distance helioseismology reveals the great potential for studying and understanding of the dynamics of the quiet Sun and active  regions, and the evolution with the solar cycle. In particular, our results show that the emergence and evolution of active regions are accompanied by multi-scale flow patterns, and that the meridional flows display the North-South asymmetry closely correlating with the magnetic activity. The latitudinal variations of the meridional circulation speed, which are probably related to the large-scale converging flows, are mostly confined in  shallow subsurface layers. Therefore, these variations do not necessarily affect the magnetic flux transport. The North-South asymmetry is also pronounced in the variations of the differential rotation ('torsional oscillations'). The calculations of a proxy of the subsurface kinetic helicity density show that the helicity does not vary during the solar cycle, and that supergranulation is a likely source of the near-surface helicity.
\end{abstract}

\section{Introduction}
\begin{figure}[t]
\begin{center}
  \includegraphics[width=\linewidth]{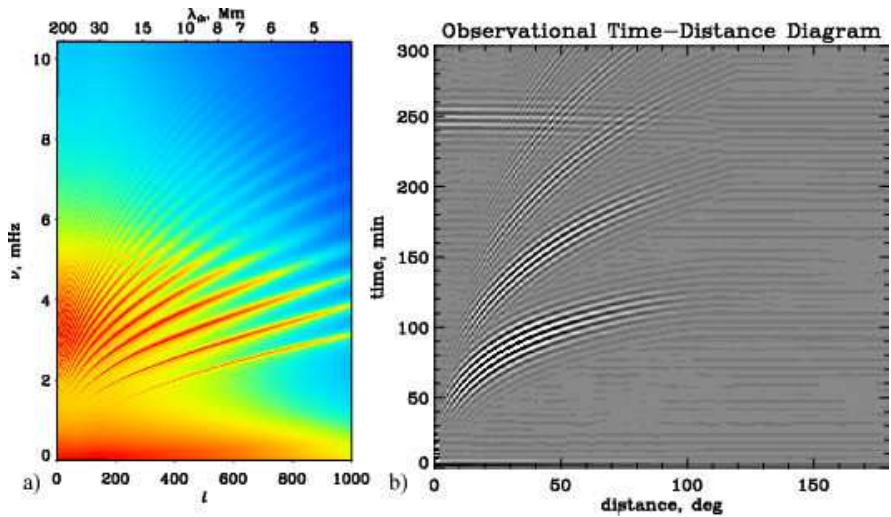}
  \caption{ a) The power spectrum of solar oscillations as a function of the angular degree $\ell$, and cyclic frequency, $\nu$. The enhanced power  corresponds to the normal oscillation modes of the Sun. b) The cross-covariance function ('time-distance diagram') of solar oscillations as a function of the distance between the correlation points on the solar surface and the time lag of the cross-covariance. The lowest ridge is formed by acoustic wave packets traveling between two surface points ('source' and 'receiver')through the solar interior (so-called, the first skip); the higher ridges are formed by the wave packets arriving to the receiver after additional reflections from the surface (the 'second' skip, and so on). 
   }\label{fig1}
\end{center}
\end{figure}

Observations of solar oscillations provide a unique opportunity to obtain information about the structure and dynamics of the solar interior beneath the visible surface. The oscillations with a characteristic period of five minutes represent acoustic waves stochastically excited by the turbulent convection in a shallow subsurface layer. The excitation mechanism has not been completely understood. However, recent numerical simulations have shown that the waves can be excited due to the interaction of turbulent vortex tubes ubiquitously generated in the intergranular lanes \citep{Kitiashvili2011}. These stochastic waves produce chaotic oscillation patterns on the solar surface. However, a spectral analysis of the time series of these patterns reveals that most of the oscillation power is concentrated in a set of normal modes (Fig.~\ref{fig1}a). These modes represent standing acoustic waves trapped in the subsurface layers by their reflection between the sharp density gradient near the surface, and the increasing sound speed in the interior. The depth of the inner reflection depends on the horizontal wavelength of the oscillations. The horizontal wavelength, $\lambda_h$, is usually represented in terms of the spherical harmonic degree, $\ell = 2\pi R/\lambda_h$. The oscillation frequency is expressed in terms of cyclic frequency $\nu=\omega/2\pi$.   In the $\ell-\nu$ diagram shown in Figure~\ref{fig1}a, the lowest ridge represents the surface gravity mode ($f$-mode). The other ridges are acoustic modes of various radial order $n$, which is equal to the number of nodes along the radius. This number is higher for higher frequency ridges. 
The time-series of solar oscillations have been obtained almost uninterruptedly since 1995 from the ground-based network GONG and space mission SOHO (Solar and Heliospheric Observatory) and SDO (Solar Dynamics Observatory). The oscillation frequencies are routinely measured from 72- and  108-day time series by fitting the modal lines which are used for inferring variations of the sound speed, asphericity, and differential rotation rate. This approach called `global helioseismology'  has provided important information about the structure, composition and dynamics of the solar interior. In particular, it was led to the discovery of a sharp radial gradient of the differential rotation at the base of the convection zone \citep{Kosovichev1996a}, the so-called tachocline, the near-surface rotational shear layer\citep{Schou1998}, subsurface zonal flows migrating with the solar activity cycle\citep{Kosovichev1997}. Recent analysis of the high-degree oscillation modes revealed a sharp gradient of the sound speed in a narrow 30-Mm deep layer just beneath the solar surface \citep{Reiter2015}. This layer (called `leptocline', \citep{Godier2001}) presumably plays an important role in the solar dynamo \citep{Pipin2011}.

It is important to note that while the oscillation power spectrum extends into the high-frequency region (10 mHz and higher), only the ridge parts with the frequency below the acoustic cut-off frequency (which is approximately at 5.2 mHz) represent the normal modes. The higher frequency parts correspond to so-called `pseudo-modes' . The pseudo-modes are formed by interference between the waves traveling from the excitation sources directly to the surface and the waves which come to the same surface location after reflection in the interior. The pseudo-mode ridges are close to the mode ridges (so that the ridges look continuous) because the excitation sources are located very close to the surface where the oscillations are observed. The pseudo-mode frequencies depend on details of the excitation mechanism and on the wave interaction with the solar atmosphere. Therefore, so-far, only the normal modes have been used for the reconstruction of solar subsurfaces. The primary restriction of global helioseismology is that it can only reconstruct the azimuthally averaged properties of the interior. This is not sufficient for the understanding of the solar dynamics and magnetism. 

The three-dimensional  structure of the solar subsurfaces can be reconstructed by techniques of local helioseismology. One of these techniques, called `ring-diagram analysis' \citep{Gough1983} is based on measuring frequency shifts in local (typically $15\times 15$ degrees) areas, and uses the global helioseismology description of the mode frequency sensitivity to local sound-speed variations and flows. This techniques allows us to reconstruct the solar subsurfaces  with relatively low spatial resolution in shallow regions. The reconstruction with higher spatial resolution and much deeper in the interior can be achieved by methods based on extracting coherent wave signals and measuring variations of the wave travel times or phase shift. These techniques called time-distance helioseismology \citep{Duvall1993} and acoustic holography \citep{Lindsey2000} employ cross-covariance functions of solar oscillations instead of the power spectral analysis. The discovery that coherent signals, such as wave packets, can be extracted from the cross-covariance functions of the stochastic solar oscillations (Fig.~\ref{fig1}b) was made by Duvall \citep{Duvall1993}. This approach was then developed  in helioseismology, terrestrial seismology, and other disciplines, and in broader applications is called `ambient noise imaging'. The foundation of this approach is based on the property of cross-covariance functions to represent wave signals corresponding to point sources. Roughly speaking the cross-covariance function can be considered as the Green's function of the solar wave equation. In  real solar conditions this is only an approximation because of the limited frequency bandwidth of solar oscillations and inhomogeneities of the solar structures and distribution of the stochastic sources. A complete theory of this approach of helioseismology has not been developed. It requires extensive studies of wave interaction with turbulence, flows and magnetic field. Nevertheless, the initial results based on relatively simple descriptions of  wave  propagation have provided important insights into the three-dimensional structures and flow patterns of the solar subsurfaces. The primary focus of these studies is mapping the flow patterns associated with the solar cycle, and formation and evolution of active regions.

In the current state of local helioseismology the systematic errors as well as effects of the stochastic realization noise have not been fully investigated. These studies require substantial effort for modeling the wave dynamics in realistic solar conditions, and require 3D MHD simulations on large supercomputer systems. The validation and testing of the time-distance technique have been performed by comparing the helioseismic inversions in the shallowest layer with the surface flows obtained by a local correlation tracking technique \citep{Liu2013}, and through the analysis and inversion of numerical simulation data for subsurface flows  and sound-speed variations \citep{Birch2011,Hartlep2013,Parchevsky2009,Parchevsky2014}. The testing for regions with strong magnetic field has not yet been completed. However, the simulations of the wave propagation in sunspot models showed that one of the primary effects in sunspot regions is the wave reflection from deeper layers, compared to the quiet-Sun regions, where the plasma parameter, $\beta=8\pi P/B^2$, the ratio of the gas pressure to magnetic pressure, is equal to unity (this layer also corresponds to the deeper photospheric surface of sunspots, known as the Wilson depression). Below the Wilson depression level the gas pressure dominates, and the helioseismic acoustic waves behave like fast MHD waves: the wave speed becomes anisotropic, and also depends on the temperature stratification beneath sunspots. The magnetic and temperature effects have not been separated in the wave-speed inversion results \citep{Kosovichev2000}. This is an important task  of local helioseismology. One of the difficulties is that the limited computer power has not allowed to simulated the sunspot models sufficiently large and deep for the helioseismology testing, so that the wave properties are not affected by the boundary conditions of the simulations. A comparison of the wave-speed inversions obtained for a sunspot regions by the time-distance and ring-diagram techniques (see \citep{Kosovichev2012,Kosovichev2011} and references therein) shows a good qualitative agreement: both inversions show a two-layer structure with a layer of reduced wave speed beneath the surface followed by a layer of an increased wave speed. However, the depth of these layers is different, perhaps, because of the drastically different spatial resolution, and different contributions of magnetic field. A comparison of the time-distance and acoustic holography inversion results has been performed by using artificial simulation data \citep{Birch2011,Parchevsky2014}.

\begin{figure}[t]
\begin{center}
  \includegraphics[width=\linewidth]{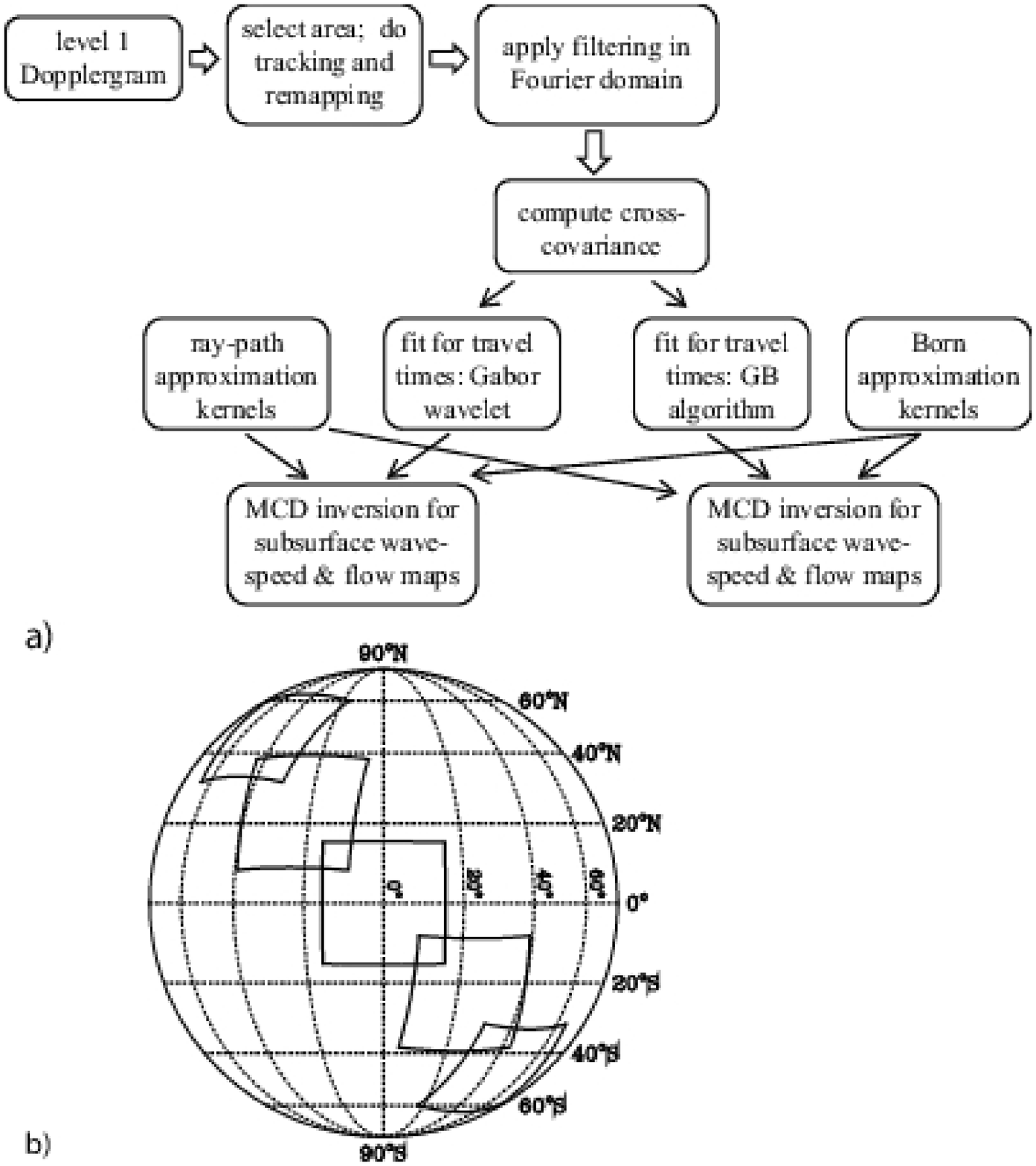}
  \caption{ a) A scheme of the Time-Distance Helioseismology Pipeline implemented at the Joint Science Operations Center (JSOC) for Solar Dynamics Observatory at Stanford University \citep{Zhao2012};  b) Illustration of the surface locations of the individual patches used for inferences of the subsurface structure and flows; the total 25 patches are used to cover $120\times 120$ degrees of the disk area.  
   }\label{fig2}
\end{center}
\end{figure}

\section{Time-Distance Helioseismology from SDO}

\begin{figure}[t]
\begin{center}
  \includegraphics[width=0.7\linewidth]{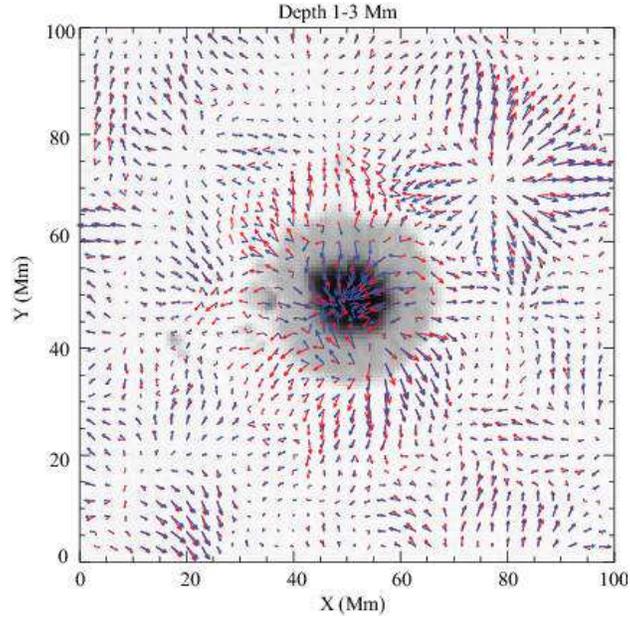}
  \caption{ Comparison of the subsurface flow maps in the depth range from 1 to 3 Mm, obtained by using two different types of the travel measurements and two different approaches for calculation of the travel-time sensitivity functions: red arrows show the flow field obtained by using the Gabor-wavelet fitting technique and the raypath kernels \citep{Kosovichev1997a}, the blue arrows are obtained by using the cross-correlation approach for the travel times \citep{Gizon2002}, and the Born-approximation kernels  \citep{Birch2000}.  The longest arrows correspond to the velocity of about 1 km/s.
   }\label{fig3}
\end{center}
\end{figure}

\begin{figure}
\begin{center}
  \includegraphics[width=0.9\linewidth]{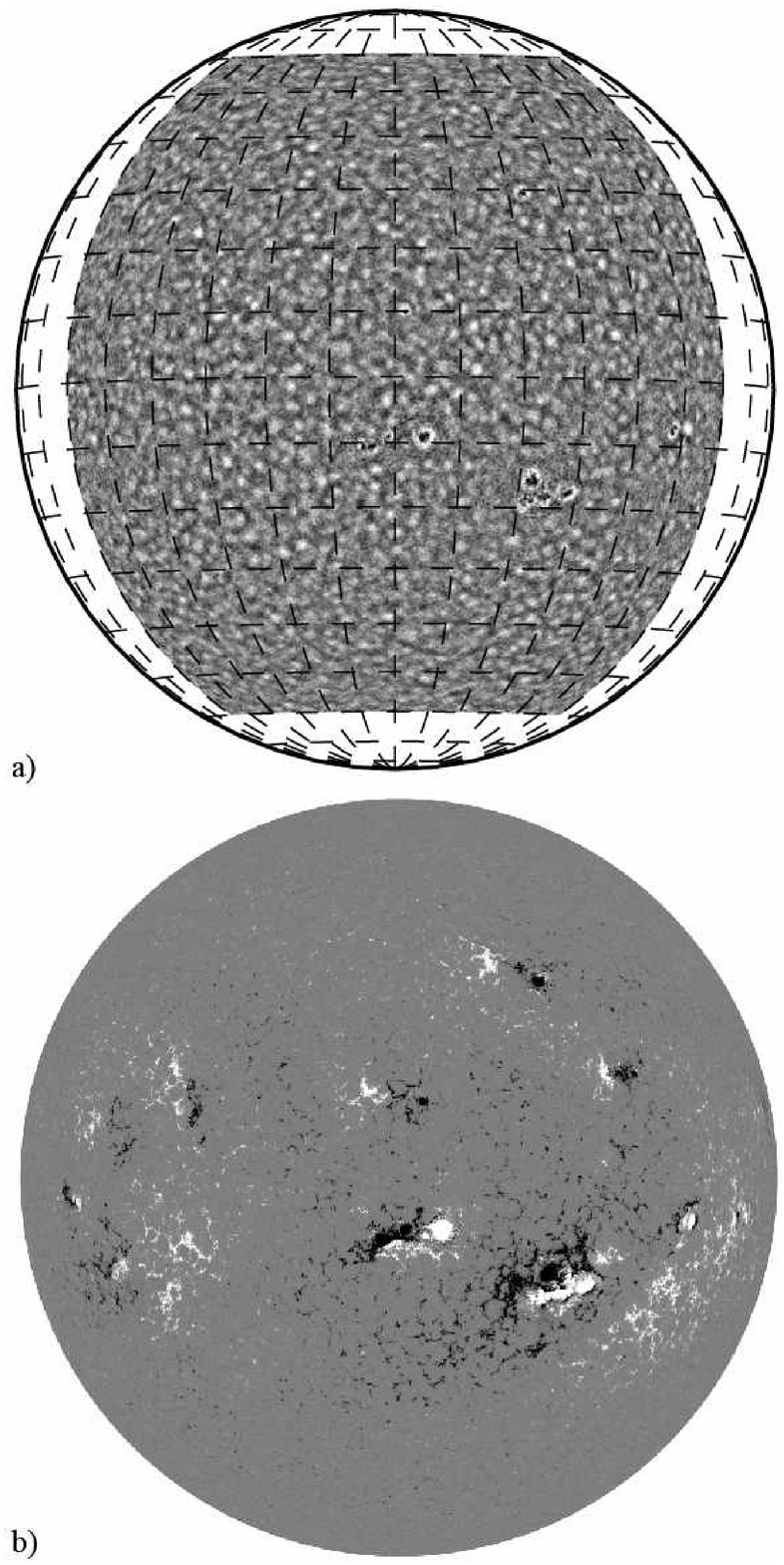}
  \caption{ a) A full-disk map showing the divergence of the horizontal velocity at the depth of 1-3~Mm, obtained on December 19, 2014, 12:00~UT. The bright point-like areas represent diverging supergranulation flows, the dark areas surrounded by bright rings represent flows converging beneath sunspots and diverging in the areas surrounding the sunspots. b) The corresponding maps of the line-of-sight magnetic field obtained from the SDO Helioseismic and Magnetic Imager.
   }\label{fig4}
\end{center}
\end{figure}

This brief review presents recent results obtained by the helioseismology reconstructions of subsurface flows in the near-surface layer and development of active regions. The results are obtained by analyzing inversions for subsurface flows from the SDO Joint Science Operations Center (JSOC) at Stanford University \citep{Scherrer2012}. The JSOC data analysis pipeline provides 3D maps of solar flows covering almost the whole disk (within 60 degrees from the disk center) in the range of depths from 0 to 30 Mm, every 8 hours. 

The time-distance helioseismology pipeline (Fig.~\ref{fig2}a) developed by the Stanford group\citep{Couvidat2012,Zhao2012} utilizes two different methods for measuring the acoustic travel times: 1) the method of fitting the Gabor wavelet to the cross-covariance function, which provides measurements of both the phase and group travel times \citep{Kosovichev1997a} and 2) the method of calculating the travel-time shift relative to a reference cross-covariance function\citep{Gizon2002}, usually calculated for a quiet-Sun region. The two sets of the travel times are calculated independently for 11 travel distances, for  the same 25 areas covering the solar disk (Fig.~\ref{fig2}b), and for the same 8-hour intervals. Then, the travel times are used for reconstruction of subsurface flows in 11 subsurface layers in the depth ranges:1 - 3, 3 - 5, 5 - 7, 7 - 10, 10 - 13, 13 - 17, 17 - 21, 21 - 26, 26 - 30, and 30 - 35 Mm,  and with the horizontal spatial sampling of 0.12 degrees (1.5~Mm) . 

\begin{figure}
\begin{center}
  \includegraphics[width=\linewidth]{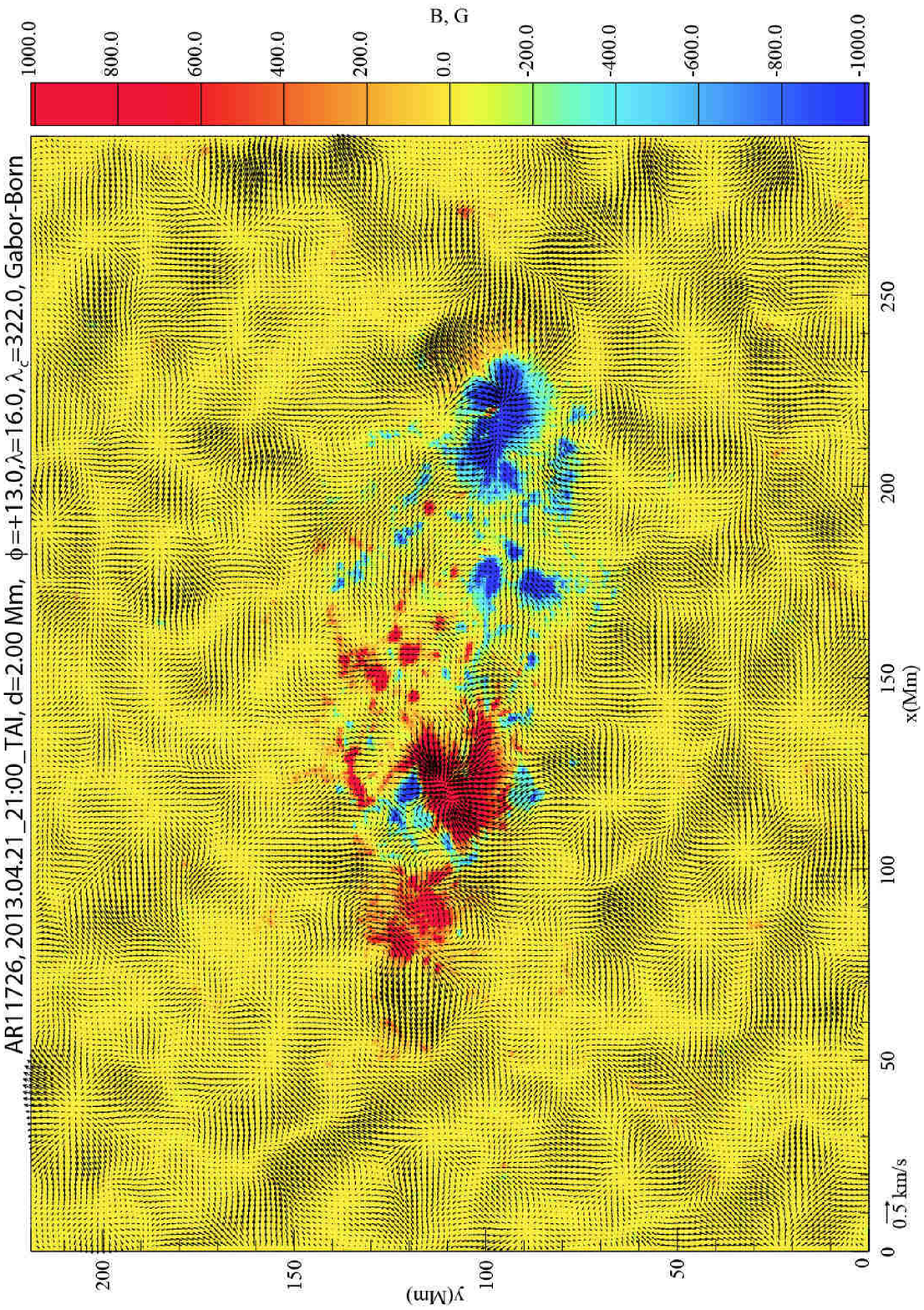}
  \caption{Arrows show the horizontal flow map in the active region NOAA 11726 at the depth of 2~Mm on April 21, 2013, 21:00~UT, about two days after its emergence from the interior. The color background image shows the surface magnetic field. The flow map is reconstructed by using the Gabor-wavelet technique for measuring the travel times, and the Born-approximation sensitivity kernels for the flow velocity inversion. 
   }\label{fig5}
\end{center}
\end{figure}

The inversions are performed by using two different methods for calculating the travel-time sensitivity functions: 1) the ray-path approximation \citep{Kosovichev1996,Kosovichev1997a} and 2) the first Born approximation \citep{Birch2000,Birch2001a,Birch2001,Birch2004,Birch2007}. The inversions are performed using the Multi-Channel Deconvolution (MCD) technique \citep{Jacobsen1999} for the two independent travel time measurements using the two types of the sensitivity kernels. Therefore, the pipeline output consists of four sets of subsurface flow maps for the same areas on the Sun \citep{Zhao2012}. This allows the comparison of the different approaches and estimate potential systematic errors. The reconstruction of subsurface flows has also been tested through analysis and inversion of numerical simulation results as well as  by the comparison of the flow maps obtained by the different techniques, and also by comparing the inversion results in the shallowest layer with the surface flows measured by the feature correlation tracking techniques \citep{Liu2013}. Figure \ref{fig3} illustrates the comparison of the flow maps below a sunspot region, obtained by using two different techniques for measuring the travel times and two different models for the travel-time sensitivity functions. The results show that the agreement is quite good everywhere except the areas close to the sunspot. As discussed in the Introduction, the effects of a strong magnetic field and large perturbations of the thermodynamic structure have not been fully investigated. Solving this problem requires more studies of systematic uncertainties using realistic numerical simulations. Nevertheless, the currently available inferences shed light on the intriguing dynamics of the solar interior.

\section{Subsurface Flows and Effects of Solar Activity}

An example of the reconstructed subsurface flow maps is illustrated in Fig.~\ref{fig4}a, which shows the distribution of the divergence of the horizontal flow velocity in the depth range 1-3~Mm. The primary feature covering the whole surface is supergranulation. The outflows of a few hundred m/s in the supergranulation cells are represented by light dot-like features. However, the examination of these maps shows that in the in the vicinity of magnetic active regions (shown in Fig.~\ref{fig4}b), the supergranulation pattern is substantially suppressed. The flow pattern beneath and around  sunspots represents a complicated combination of converging flows towards the sunspot centers (displayed as dark dots in the divergence map) surrounded by outflows (represented as white rings). Such a pattern, similar to the flows shown in Figures~\ref{fig3} and \ref{fig5}, has been previously studied using SOHO/MDI and Hinode data \citep{Zhao2001,Zhao2009b,Zhao2010}. A new feature of the SDO/HMI analysis is that the HMI data allow us to reconstruct the flows in a shallow subsurface layer, and match these to the directly observed surface flows. This agreement provides more confidence in the helioseismic inferences.

Figure~\ref{fig5} shows a portion of the horizontal velocity map around an emerging active region NOAA 11726, during its development phase. This is the largest active region observed by the HMI instrument during the first five years of operation. The flow velocities are shown by arrows, and the photospheric magnetogram is represented by the color map. Such flow maps are obtained with one-hour sampling, although each map requires eight-hour time series of Dopplergrams for the helioseismology analysis. The analysis of these maps indicates that the converging flows beneath the sunspots are developed simultaneously with the sunspot formation, as was previously found from analysis of the SOHO/MDI data \citep{Kosovichev2006,Kosovichev2009}, and, probably, is closely associated with the mechanism of the sunspot formation. At the same time, large-scale diverging flows are developed around the active region, and are probably related to the well-known phenomenon of the surface `moat' flow. 

\begin{figure}
\begin{center}
  \includegraphics[width=\linewidth]{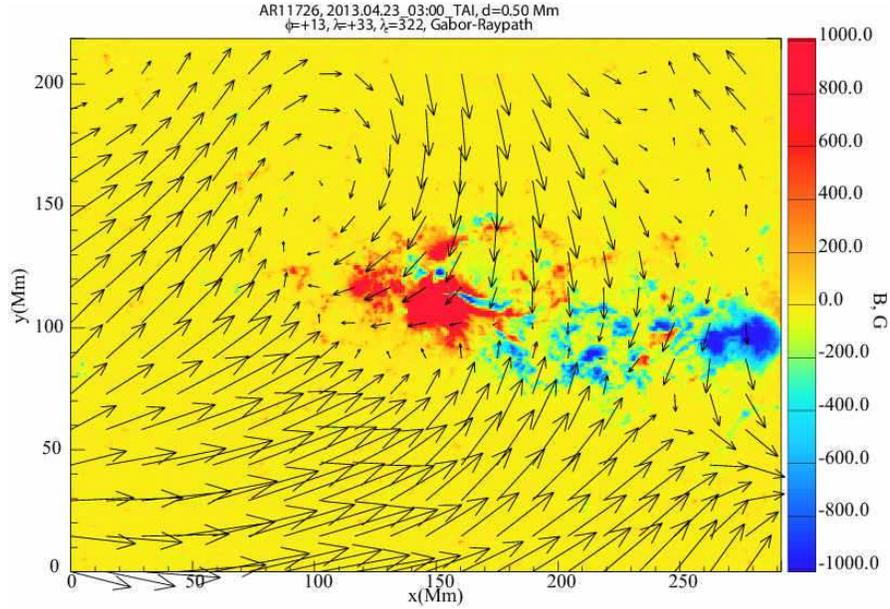}
  \caption{Arrows show the horizontal flow map around the active region NOAA 11726 at the depth of 0.3~Mm on April 23, 2013, 03:00~UT (when the active region is fully developed) after averaging the high-resolution flow map on a grid with a 15-degree sampling. The color background image shows the surface magnetic field. The typical flow speed is about 50~m/s.
   }\label{fig6}
\end{center}
\end{figure}

Outside the active region the flow pattern is mostly represented by supergranulation which, however, is clearly disturbed by the presence of the active region. It is interesting that the spatial averaging of these flow maps reveals a large-scale pattern of converging flows occupying a surrounding  area which is significantly larger than the active region (Fig.~\ref{fig6}). Such converging flows with the characteristic speed of about 50~m/s were first discovered by the ring-diagram technique \citep{Haber2003}. The origin of these flows is not understood, but our analysis shows that these flows are formed and stable only when the active region is fully developed, and, thus, they are not associated with the emergence of magnetic flux and formation of the active region. 

\begin{figure}
\begin{center}
  \includegraphics[width=\linewidth]{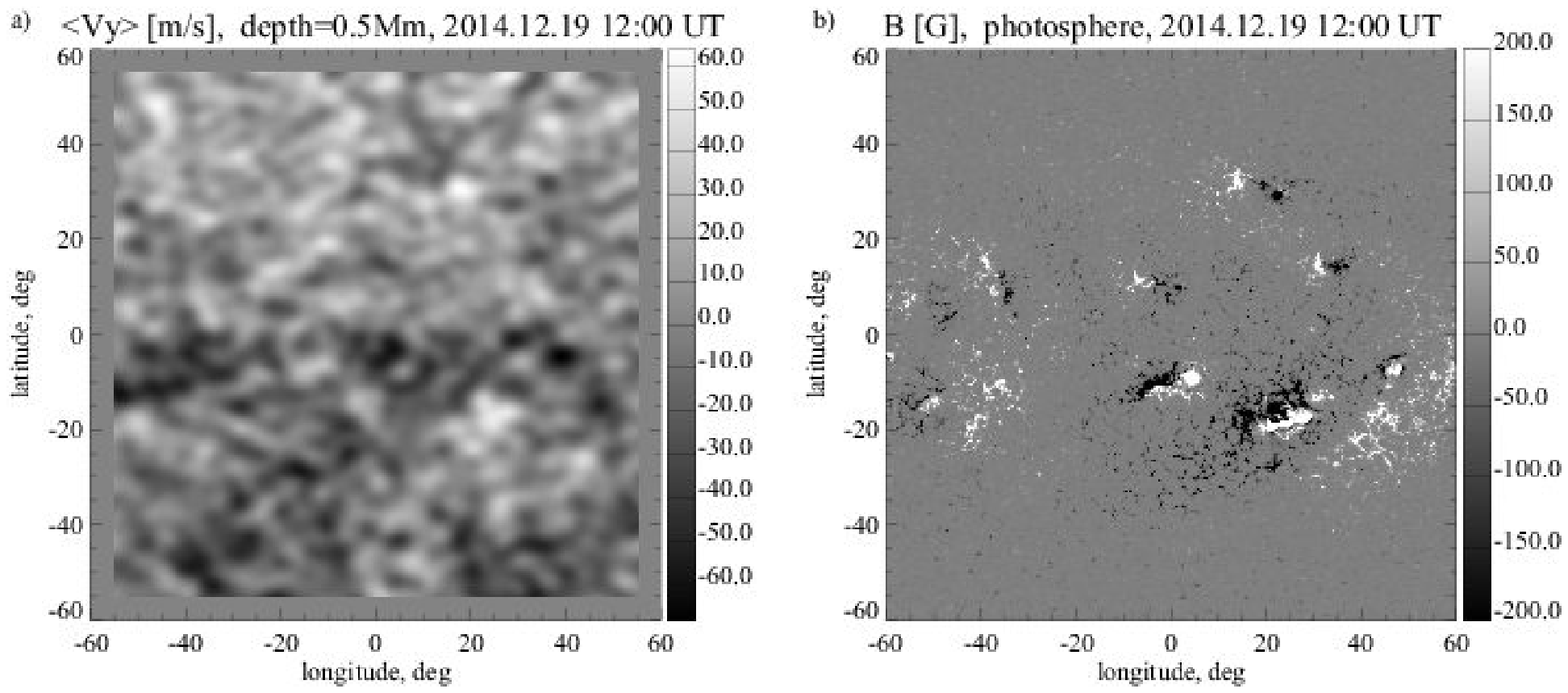}
  \caption{ a) A map of the North-South component of the subsurface flow velocity of December 19, 2014, 12:00~UT, smoothed with a $5\time 5$~deg Gaussian window, reveals that the meridional flow pattern disturbed the converging flows around active regions. b) The corresponding photospheric magnetogram map. Red color shows the positive polarity, the blue color shows the negative polarity. 
   }\label{fig7}
\end{center}
\begin{center}
  \includegraphics[width=\linewidth]{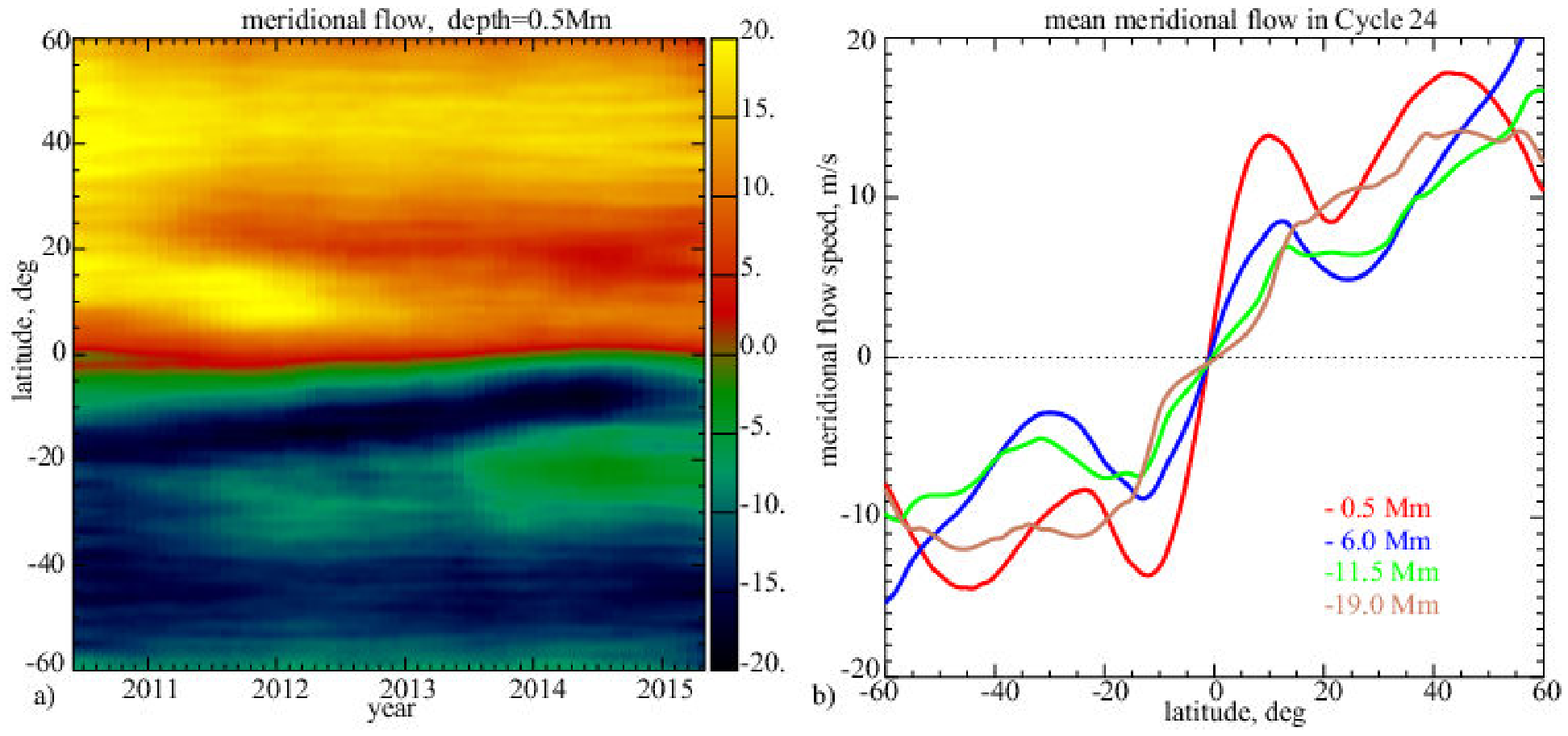}
  \caption{a) Evolution of the subsurface meridional flows obtained from the 5-years of the SDO/HMI observations during Solar Cycle 24. The red and yellow colors show the flow components towards the North pole, the green and blue colors show the South-ward flow. The color scale range is from -20 to 20 m/s. b) The mean meridional flow averaged for the whole period of observations at four different depths.  
   }\label{fig8}
\end{center}
\end{figure}
The large-scale converging flows around active regions play an important role in the solar-cycle evolution of the meridional circulation \citep{Haber2002,Zhao2004}. The meridional circulation can be calculated from the reconstructed subsurface flow maps by averaging the North-South component of the flow velocity. Figure~\ref{fig7}a shows a map of the North-South velocity component smoothed with a 5-degree Gaussian window. 

\begin{figure}
\begin{center}
  \includegraphics[width=\linewidth]{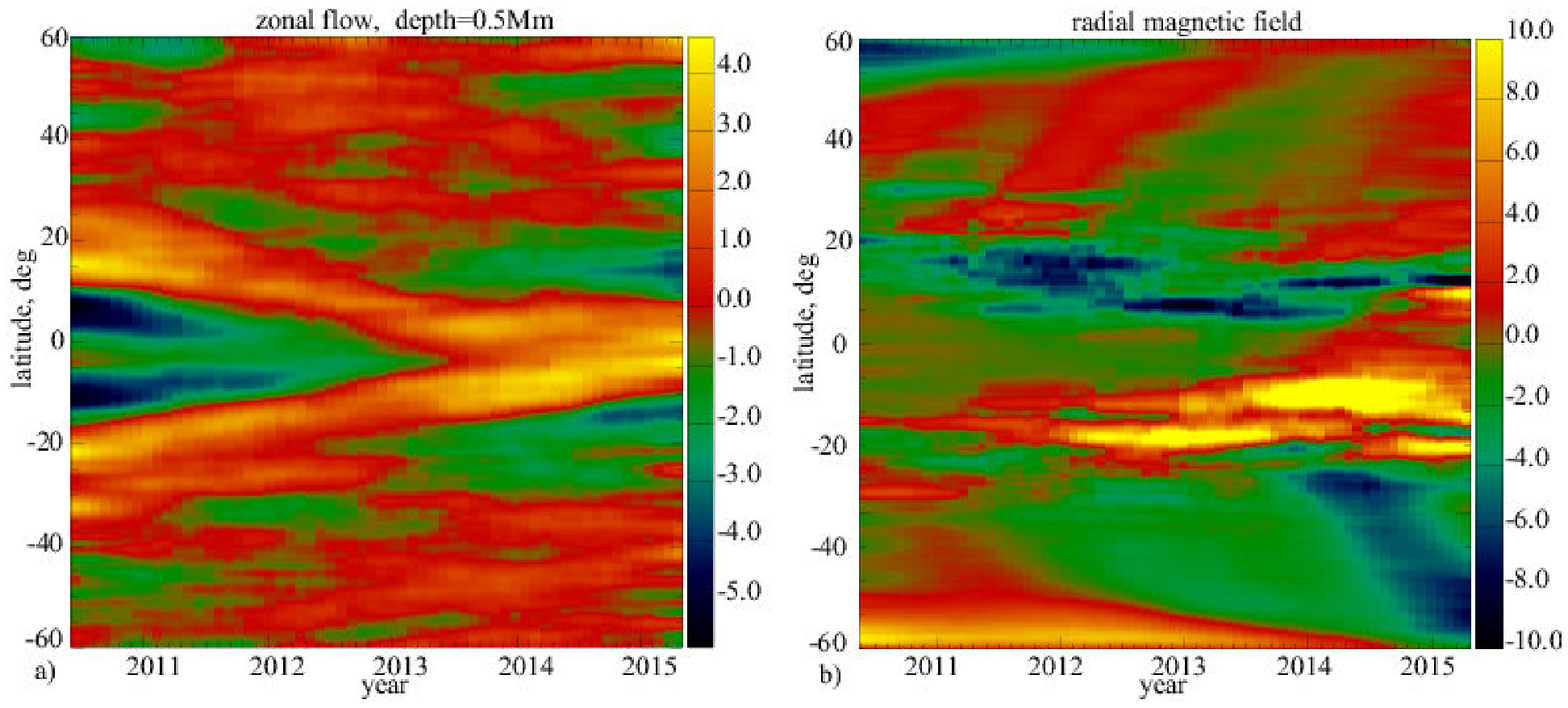}
  \caption{a) The evolution of the subsurface zonal flows (`torsional oscillation') during the first five years of the SDO observations, covering the raising phase of Solar Cycle 24. The flow map is obtained by averaging of the azimuthal flow component of longitude and one-month time, subtracting the mean rotational velocity from each of the one-month averaged profile, and then stacking the residual velocity profiles and smoothing with one-year running window to remove the annual variations due to the inclination of the Earth orbit. The yellow and red colors correspond to the zonal flows faster than the mean solar rotation at the same depth, and the blue color shows the slower rotating regions. b) The corresponding magnetic `butterfly' diagram from the SDO/HMI data, showing the evolution of the mean radial magnetic field in the solar photosphere during the same period.   
   }\label{fig9}
\end{center}
\end{figure}

The appearance of the poleward trends in each hemisphere is apparent. It is interesting that the meridional circulation can be detected in a single flow map, but it is also important that the flow pattern correlates with the surface magnetic field map shown in Fig.~\ref{fig7}b. The variations can be interpreted as caused by the  large-scale converging flows around active regions. However, the strongest variations in this flow map seem to be in the areas of decaying active regions. Thus, it is important to investigate the formation and evolution of the converging flows during the whole evolution of active regions, from their formation to decay.

The evolution of meridional circulation is obtained by averaging the North-South velocity component over longitude and one-month periods, and displaying the averages in the form of a time-latitude diagram (Fig.~\ref{fig8}a). This diagram shows that the evolution of the subsurface meridional circulation correlates with the magnetic activity in each hemisphere. At the beginning of the current cycle most active regions emerged in the Northern hemisphere, where we see a strong variation of the meridional circulation speed: a sharp increase at low latitudes (in the 10-20 deg interval) and a decrease at mid latitudes (in the 20-30 deg range). A similar variation in the Southern hemisphere is observed in 2014-15 when most magnetic activity was in the South (Fig.~\ref{fig9}b). 

Such variations of the meridional circulation may affect the magnetic flux transport and the polar  magnetic field polarity reversal. However, this link has not been fully established \citep{vSvanda2007,vSvanda2007a,vSvanda2008}. Figure~\ref{fig8}b shows the variation of the mean (averaged over the whole period) meridional circulation profile with depth. It appears that at a depth of $\sim 10$~Mm the latitudinal variations are significantly reduced, and at the depth of $\sim 20$~Mm almost entirely disappear. Therefore, if the large-scale magnetic flux is anchored at this depth or lower then its transport is not affected by the meridional flow variations.

Solar-cycle variations of the differential rotation are also of great interest for the understanding of the mechanisms of solar activity. These variations, known as `torsional oscillation', have been detected from the surface Doppler-shift maps \citep{Howard1980,Ulrich2001}, and by global \citep{Kosovichev1997} and local \citep{Zhao2004} helioseismology. The high-resolution flow maps from SDO/HMI provide new opportunities for investigating the detailed structure and evolution of these flows. Figure~\ref{fig9}a shows the time-latitude diagram of variations of the differential rotation during the five years of the SDO/HMI observations. These variations are relative to the mean differential rotation profile averaged for the whole period, and smoothed with a one-year window to remove the orbital variations. As it was established before, the zonal flow closely correlates with the magnetic butterfly diagram (Fig.~\ref{fig9}b). However, our results also show the North-South asymmetry of the flows, which follows the asymmetry of the magnetic activity. 

\begin{figure}[t]
\begin{center}
  \includegraphics[width=\linewidth]{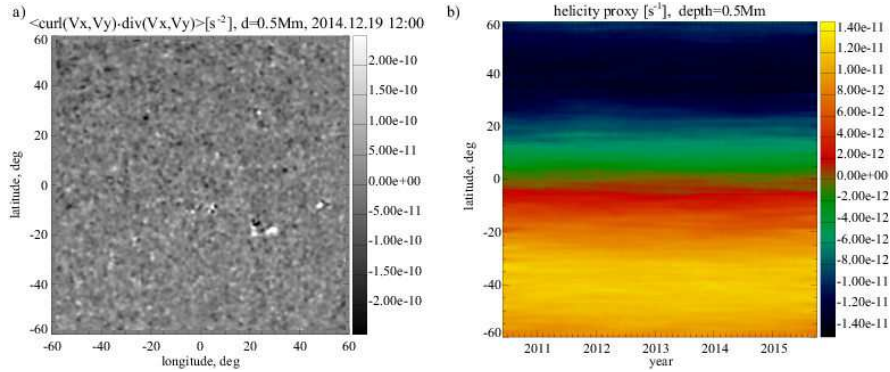}
  \caption{ a) The proxy of kinetic helicity density,  ${\nabla}{\vecv_h}\cdot({\nabla}\times{\vecv_h})_z$,
calculated from the flow map of December 19, 2014, 12:00~UT, reveals a systematic North-South asymmetry in local sources associated with supergranulation. b) The evolution of the mean helicity proxy during the observed period of Solar Cycle 24. 
   }\label{fig10}
\end{center}
\end{figure}

The flows maps allow us to investigate other important properties of the subsurface dynamics of the Sun, which previously were  not accessible. For illustration, in Figure~\ref{fig10}a we show a map of the kinetic helicity proxy calculated as    ${\nabla}{\vecv_h}\cdot({\nabla}\times{\vecv_h})_z$, where $\vecv_h$ is the horizontal velocity component. By looking at this map one can  notice that the Northern hemisphere is darker than the Southern hemisphere, and that the asymmetry is particularly pronounced in the supergranulation cells. After the longitudinal and time averaging of the individual helicity proxy maps we obtain the time-latitude diagram (Fig.~\ref{fig10}b), which shows that the kinetic helicity does not vary on this time scale. This result puts constraints on the dynamo theories, and also shows that the supergranulation flows are likely a primary source of the near-surface helicity.

\section{Conclusion}
The initial analysis of the subsurface flow maps reconstructed from the five years of  SDO/HMI data by time-distance helioseismology reveals the great potential for studying and understanding the dynamics of the quiet-Sun and active  regions, and the evolution with the solar cycle. In particular, our results show that the emergence and evolution of active regions are accompanied by multi-scale flow patterns. Beneath the sunspot, during their  formation, we observe appearance of flows converging towards the sunspot center. and also the 'moat'-like flows diverging from the active region in the surrounding regions. On the larger scale, revealed by averaging the high-resolution flow maps, we find a pattern of flows converging towards the active region. This pattern is formed when the active region is fully developed.  On the global-Sun scale, the flow maps allow us to investigate the structure and evolution of the meridional flows. In particular, we find that the meridional flows display the North-South asymmetry closely correlating with the magnetic activity. The latitudinal variations of the meridional circulation speed, which are probably related to the large-scale converging flows, are mostly confined in a shallow subsurface layers. Therefore, these variations do not necessarily affect the magnetic flux transport. The North-South asymmetry is also pronounced in the variations of the differential rotation ('torsional oscillation'). The calculations of a proxy of the subsurface kinetic helicity density show that the helicity does not vary during the solar cycle, and that the supergranulation is a likely source of the near-surface helicity. These initial results are obtained from the analysis of a small sample  of flow maps produced by the SDO/HMI time-distance helioseismology pipeline. Further detailed investigations are required for understanding the complicated subsurface dynamics of the Sun.

\section*{Acknowledgment}
This work was supported by the CNRS, and NASA grants NNX09AJ85G and
NNX14AB70G .

\newpage
\addcontentsline{toc}{section}{References}
\bibliographystyle{spbasic}

\end{document}